\documentclass[12pt,a4paper,superscriptaddress,preprint]{revtex4}

\usepackage[dvips]{graphicx}
\usepackage{amssymb,amsmath}
\usepackage{color}
\usepackage{bm}
\usepackage{booktabs}

\usepackage{color}

\newcommand{\green}[1]{}
\newcommand{\blue}[1]{}

\begin{document}

\newcommand{\EQ}{Eq.~}
\newcommand{\EQS}{Eqs.~}
\newcommand{\FIG}{Fig.~}
\newcommand{\FIGS}{Figs.~}
\newcommand{\TAB}{Tab.~}
\newcommand{\TABS}{Tabs.~}
\newcommand{\SEC}{Sec.~}
\newcommand{\SECS}{Secs.~}

\title{Enhancing the spectral gap of networks by node removal}

\author{Takamitsu Watanabe}
\affiliation{Department of Physiology, School of Medicine,
The University of Tokyo,
7-3-1 Hongo, Bunkyo-ku, Tokyo 113-8656, Japan}
\author{Naoki Masuda\footnote{Corresponding author: masuda@mist.i.u-tokyo.ac.jp}\ }
\affiliation{Graduate School of Information Science and Technology,
The University of Tokyo,
7-3-1 Hongo, Bunkyo-ku, Tokyo 113-8656, Japan}
\affiliation{PRESTO, Japan Science and Technology Agency,
4-1-8 Honcho, Kawaguchi, Saitama 332-0012, Japan}

\begin{abstract}
Dynamics on networks are often characterized by the second smallest eigenvalue of the Laplacian matrix of the network, which is called the spectral gap.  Examples include the threshold coupling strength for synchronization and the relaxation time of a random walk. A large spectral gap is usually associated with high network performance, such as facilitated synchronization and rapid convergence. In this study, we seek to enhance the spectral gap of undirected and unweighted networks by removing nodes because, practically, the removal of nodes often costs less than the addition of nodes, addition of links, and rewiring of links.  In particular, we develop a perturbative method to achieve this goal. The proposed method realizes better performance than other heuristic methods on various model and real networks. The spectral gap increases as we remove up to half the nodes in most of these networks.
\end{abstract}

\pacs{89.75.Fb, 64.60.aq, 05.45.Xt}

\maketitle

\section{Introduction}
Various systems of interacting elements can be represented by networks that consist of a set of nodes and links that connect pairs of nodes. The structure of networks affects various dynamics occurring on the networks \cite{Pikovsky01book, Boccaletti06PR, Barrat08book}.In particular, many dynamics on networks are controlled by a few extremal eigenvalues of the adjacency matrix and the Laplacian matrix of the network  (see \SEC\ref{sec:methods} for the definition of the Laplacian matrix). The values of these eigenvalues provide concise and useful information about the dynamics on the networks.

In this study, we focus on the second smallest eigenvalue of the Laplacian matrix; it is called the spectral gap and is denoted by $\lambda_2$ in this paper.  We examine $\lambda_2$ because it characterizes a wide class of dynamics on networks as follows.
First, a network with a large value of $\lambda_2$ decreases the threshold of coupling strength for synchronization for both linear dynamics and some nonlinear dynamics including coupled oscillators on networks \cite{Pikovsky01book, Boccaletti06PR, Barrat08book, Almendral07new, Arenas08pr, Motter07new, Donetti06jstatm}. 
Such a family of nonlinear dynamics is called the class II 
\cite{Boccaletti06PR, Almendral09pre} or type II \cite{Arenas08pr} dynamics.
Note that the largest Laplacian eigenvalue as well as $\lambda_2$
is an important determinant of the synchronizability in the so-called
class III
\cite{Boccaletti06PR,Almendral09pre} or type I \cite{Arenas08pr} dynamics.
However, we are not concerned with class III or type I dynamics in this paper.
Second, when $\lambda_2$ is large, synchronization in these dynamics \cite{Almendral07new} and consensus dynamics \cite{Olfati07} occur rapidly in certain types of networks.
Third, $\lambda_2$ characterizes the convergence speed of the Markov chain on the network to the stationary density 
\cite{Donetti06jstatm, Cvetkovic10book}.
Fourth, the first-passage time of the random walk is characterized by $\lambda_2$ \cite{Donetti06jstatm}.
Fifth, the duality between the coalescing random walk and the voter model \cite{Liggett85book-Durrett88book}implies that $\lambda_2$ also determines the consensus time of the stochastic voter dynamics. This is in agreement with the results obtained for the majority-vote spin dynamics on networks \cite{Almendral07new}.
In addition to these dynamical properties of networks, various graph-theoretical structural properties of networks are characterized by $\lambda_2$ \cite{Mohar91book,Cvetkovic10book}.

In these applications, a large value of $\lambda_2$ is usually preferred because it indicates, for example, enhanced synchronizability and fast convergence. Consequently, the enhancement of $\lambda_2$ has been explored in the framework of designing of networks \cite{Motter07new, Nishikawa06PRE_PhyD} and numerical optimization via the rewiring of links \cite{Donetti05prl, Donetti06jstatm}.
In practice, however, rewiring links, constructing optimized networks from scratch, and adding nodes or links are likely to cost more than the removal of nodes or links of a given network.
The effects of removal of nodes or links have been investigated in the context of the cascading failure \cite{Motter04prl} and the influence on extreme eigenvalues of the adjacency matrix \cite{Restrepo06prl}.
With regard to Laplacian eigenvalues, the removal of links always decreases $\lambda_2$ 
and makes the network less likely to synchronize
\cite{Milanese10pre,  Nishikawa2010PNAS}.
However, to the best of our knowledge, whether or not careful removal of  nodes may increase $\lambda_2$ has not yet been examined. We treat this problem in the present paper.
                
Although removal of nodes generally decreases the magnitude of
activities, stabilizing
synchronization at the expense of the magnitude
is valuable in some applications.  The
treatment of cardiac arrhythmia is one of the examples. The heart consists
of a large number of cardiac cells that show nonlinear dynamics
\cite{Guevara1981Science_Chialvo1987Nature}. Synchronized
dynamics of cardiac cells create physiological heart beats
\cite{Bub1998PNAS_GuytonBook}. Cardiac arrhythmia is considered to
be caused by malfunction of synchronization.
The catheter ablation aims at restoring synchrony of the entire heart by
electrically deactivating some cardiac cells that prevent
synchronization \cite{Glass01nat}.  
As another example, proper operations of
power plant networks also critically require
that the frequency of voltage among power
plants is synchronized
 \cite{Filatrella2008EPJ, Fioriti2009CIIS}. Loss of
the synchronization may induce a blackout in the entire network.
Therefore, it is likely that stable synchrony at the expense of some
total power supply serves steady supplying of electricity
to the entire network
\cite{Filatrella2008EPJ}.

We compare various strategies for maximizing $\lambda_2$ by sequential node removal on various model and real networks. In particular, we develop a perturbative strategy that is applicable to relatively large networks in terms of the computational cost. We show that the performance of the perturbative strategy is comparable to that of the computationally costly optimal sequential
strategy and is generally better than that of heuristic strategies.
In addition, in many examined examples, $\lambda_2$ continues to increase until we remove a fairly large fraction of nodes ($\approx$ 50\%) sequentially according to the perturbative strategy.

\section{Strategies for sequential node removal}\label{sec:methods}

We consider undirected and unweighted connected networks with $N$ nodes.
The Laplacian matrix $L$ is defined as follows. 
$L_{ij}$ ($1\le i\neq j\le N$)
is equal to $-1$ if node $i$ and $j$ are connected and 0 otherwise;
$L$ is a symmetric matrix. The diagonal is given by $L_{ii} = k_i$, where $k_i$ is the degree of node $i$. Note that $\sum_{j=1}^N L_{ij}=0$ for each $i$.
$L$ has $N$ (real) nonnegative eigenvalues $0=\lambda_1<\lambda_2\le \cdots \le \lambda_N$. We seek to maximize $\lambda_2$ upon sequential removal of nodes. We compare the effectiveness of the following node removal strategies by applying them to model and real networks.

\begin{itemize}

\item \textit{Degree-based strategy}: In each step, we remove the node with the smallest degree in the remaining network. The rationale behind this strategy is that the smallest degree controls $\lambda_2$, with a useful bound being 
$\lambda_2\le k_{\min}N/(N-1)$, where $k_{\min}$ is the smallest degree in the network \cite{Arenas08pr,Mohar91book,Cvetkovic10book}. If there exist multiple nodes having the same smallest degree, we select one of them with an equal probability.

In intentional attacks on networks, where the aim is to fragment the network into disjoint components with a small number of removed nodes, removing nodes with the largest degree is an effective strategy \cite{Albert00nat-Callaway00prl-Cohen01prl}.  We implemented this strategy but obtained poor results for our purpose, and therefore we do not mention it in the following.

\item \textit{Betweenness-based strategy}:
In each step, we remove the node with the smallest betweenness centrality. The betweenness centrality of node $i$ is defined as follows. 
Denote by $\sigma_{i_1i_2}$ the number of the shortest paths between nodes
$i_1$ and $i_2$, and by $\sigma_{i_1i_2}(i)$ 
the number of the shortest paths between them that
pass through node $i$. We set $\sigma_{i_1i_2}(i_1)=\sigma_{i_1i_2}(i_2)=0$.
The betweenness centrality of node $i$ is proportional to
$\sum_{i_1=1; i_1\neq i}^N \sum_{i_2=i_1+1; i_2 \neq i}^N 
\sigma_{i_1i_2}(i) / \sigma_{i_1i_2}$ \cite{Freeman79,Boccaletti06PR, Barrat08book}.
Sequentially removing nodes with the largest betweenness centrality yielded poor results, and therefore we do not mention it in the following.

\item \textit{Optimal sequential strategy}:
We calculate the change in $\lambda_2$ induced by the removal of each node by direct numerical simulations. Then, we select the node whose removal increases $\lambda_2$ by the largest amount.
Note that this strategy is computationally costly because it requires the calculation of $\lambda_2$ for $N$ different networks, each having $N-1$ nodes. Calculating $\lambda_2$ for a single network requires $\mathrm{O}(N^3)$ time. Therefore, carrying out a single step of the optimal sequential strategy requires $\mathrm{O}(N^4)$ time.

\item \textit{Perturbative strategy}:
To avoid the computational cost of the optimal sequential strategy, we develop an approximate perturbative strategy defined as follows. 
Related perturbative calculations are treated in \cite{Restrepo06prl,Masuda09njp,Milanese10pre}.

Let us represent the eigen equation for $\lambda_2$ as $L\bm u=\lambda_2 \bm u$, where $\bm u$ is the $N$-dimensional eigenvector of $L$ corresponding to $\lambda_2$. The eigenvector $\bm u$ is normalized such that $\sum_{i=1}^N u_i^2=1$, where $u_i$ is the $i$th element of $\bm u$.
The eigen equation after the removal of node $i$ is given by
\begin{equation}
	(L + \Delta L)(\bm u + \Delta \bm u) = (\lambda_2 + \Delta \lambda_2)(\bm u + \Delta \bm u),
	\label{eq:def_eigenvalue}
\end{equation}
where the changes in $L$, $\lambda_2$, and $\bm u$ induced by the removal of node $i$ are denoted by $\Delta L$, $\Delta \lambda_2$, and $\Delta \bm u$, respectively.
Because $L$ is symmetric, the displacement matrix $\Delta L$ is given by 
$(\Delta L)_{ii} = -L_{ii}=-k_i$,
$(\Delta L)_{jj} = L_{ji} (j \neq i)$, and
$(\Delta L)_{ji} = (\Delta L)_{ij} = -L_{ji} (j \neq i)$.
Because the $i$th component of $\bm u + \Delta \bm u$ is equal to zero, we write 
$\Delta \bm u = \delta \bm u - u_i \hat{\bm e}_i$, 
where $\hat{e_i}$ is the unit vector for the $i$th component and $\delta \bm u$ is an $N$-dimensional vector. 
By multiplying the normalized left eigenvalue $\bm u^{\top}$ ($\top$ denotes the transpose) from the left of 
Eq.~\eqref{eq:def_eigenvalue},
we obtain 
\begin{equation}
	\Delta \lambda_2 = 
	\frac{\bm u^{\top} \Delta L (\bm u - u_i \hat{\bm e}_i + \delta \bm u)}{\bm u^{\top}(\bm u - u_i \hat{\bm e}_i + \delta \bm u)}
	\label{eq:def_before_approximation}
\end{equation}
We assume that the absolute value of each element of $\delta \bm u$
is smaller than that of $\bm u - u_i \hat{\bm e}_i$. Then,
by ignoring $\delta\bm u$ in Eq.~\eqref{eq:def_before_approximation}
and substituting the expression for $\Delta L$ in Eq.~\eqref{eq:def_before_approximation}, we obtain
\begin{equation}
	\Delta \lambda_2 \approx
	\frac{\sum_{j \in {\mathcal N}_i} u_j(u_i - u_j)}{1-u_{i}^{2}},
	\label{eq:perturbative_result}
\end{equation}
where ${\mathcal N}_i$ indicates the neighborhood of
node $i$.

In the perturbative strategy, we remove node $i$ that maximizes $\Delta \lambda_2$ given by Eq.~\eqref{eq:perturbative_result}. Note that carrying out one step of the perturbative strategy requires solving the eigen equation just once. Therefore, the computation cost is $\mathrm{O}(N^3)$, which is $N$ times smaller than that for the optimal sequential strategy.
In the following numerical simulations,
the networks are connected during sequential node removal
for all the networks and strategies.

\end{itemize}

\section{Results}

In this section, we apply the node-removal strategies introduced in Sec.~\ref{sec:methods} to various model and real networks.

\subsection{Model networks}

First, we apply different strategies to the following types of model networks.

\begin{itemize}

\item Erd\H{o}s-R\'{e}nyi (ER) random graph with mean degree $\left<k\right>=p(N-1)$, where $p$ is the probability that a link exists between a pair of nodes.

\item Watts-Strogatz (WS) model \cite{Watts98nat}, where each node is connected to $\left<k\right>/2$ closest nodes on each side along the ring and a fraction, 0.3, of links are rewired randomly.

\item Barab\'{a}si-Albert (BA) model \cite{Barabasi99sci}, a representative growing scale-free network model. We start the growth of the network from the complete graph of $m$ nodes and add a node with $m$ links one-by-one according to the preferential attachment. 
We obtain $\left<k\right>\approx 2m$, degree distribution $p(k)\propto k^{-3}$, and low clustering.

\item Holme-Kim (HK) model \cite{Holme02pre}, 
a growing scale-free network model. The algorithm of the HK model is similar to that of the BA model. The difference is that, when a node is added, the preferential attachment is used with a certain probability, which we set as 0.5. With the remaining probability (\textit{i.e.}, 0.5), we use the so-called triad formulation rule to enhance clustering. We obtain $\left<k\right>\approx 2m$, degree distribution $p(k)\propto k^{-3}$, and high clustering.

\item Goh model \cite{Goh01prl}, 
a nongrowing scale-free network model. We assign the weight $w_i=i^{-0.5}$ to each node $i$. Then, we select a pair of nodes with the probability proportional to $w_i$ and connect them. We repeat this procedure until we obtain the desired mean degree $\left<k\right>$. We obtain $p(k)\propto k^{-3}$.

\end{itemize}
For each network model, we assume two values of $\left<k\right>$. For each case, we carry out sequential node removal according to different strategies. Because, 
in stepwise node removal, the optimal sequential strategy is usually an efficient way, we will mainly evaluate the effectiveness of the other strategies using the performance of the optimal sequential strategy as a baseline.

\begin{figure}
\begin{center}
\includegraphics[width=15cm]{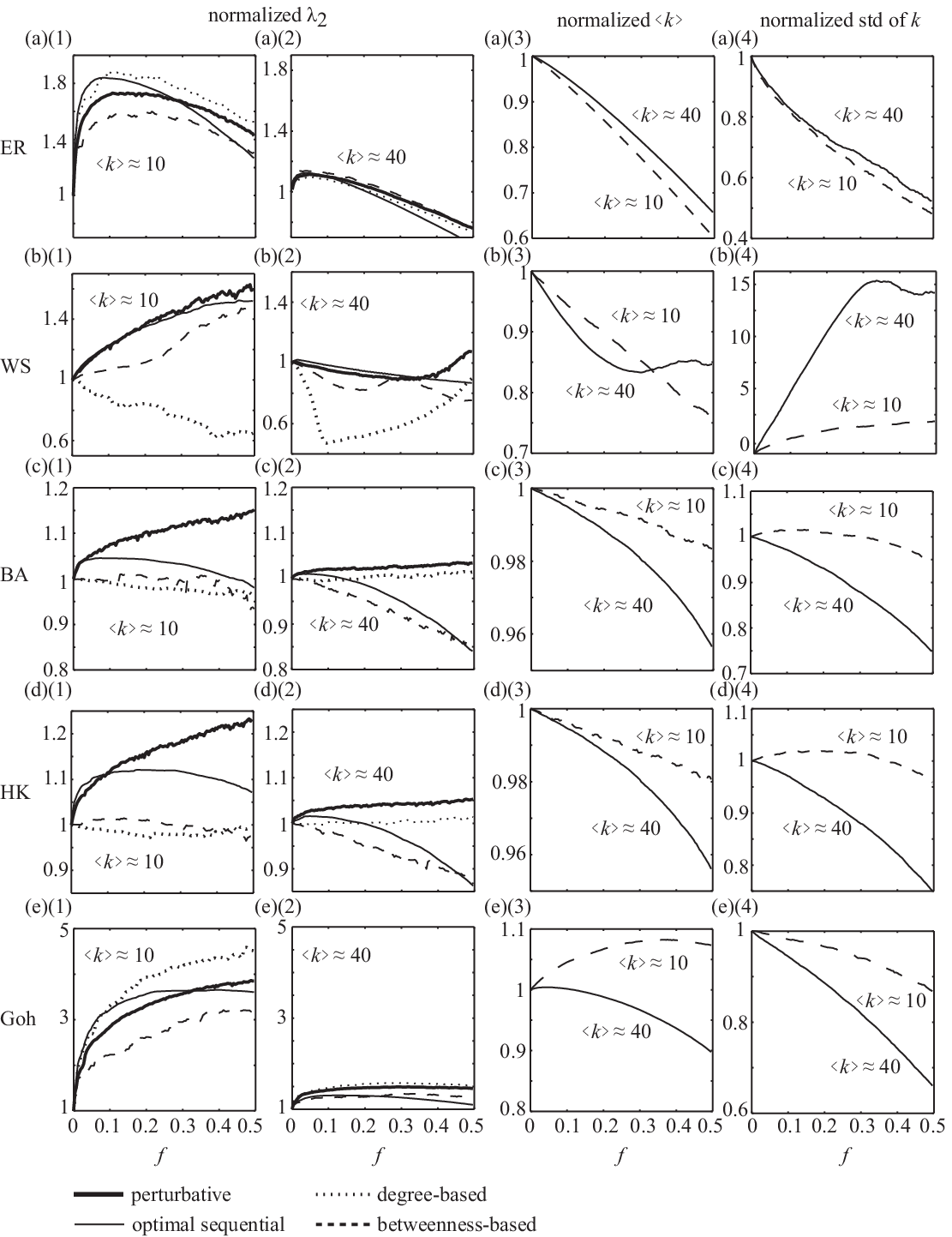}
\caption{Numerical results for model networks with
$N=250$ nodes. We set $\left<k\right>=10$ and $\left<k\right>=40$ for each network.
(a) ER random graph,
(b) WS model,
(c) BA model, 
(d) HK model, 
and (e) Goh model.
(a)(1), (a)(2), (b)(1), (b)(2), (c)(1), (c)(2), (d)(1), (d)(2), (e)(1), and
(e)(2) show the change in 
$\lambda_2$ induced by each strategy.
The mean degrees 
are shown in the panels.
(a)(3), (b)(3), (c)(3), (d)(3), and (e)(3) show 
the change in the mean degree with node removal.
(a)(4), (b)(4), (c)(4), (d)(4), and (e)(4) show
the change in the standard deviation of the degree with
node removal.}
\label{fig:1}
\end{center}
\end{figure}

The numerical results obtained for the networks with $N=250$ averaged over 10 trials are shown in \FIG\ref{fig:1}. Figures~\ref{fig:1}(a)(1) and 1(a)(2) show the values of $\lambda_2$ after removing a fraction of nodes for two ER random graphs with different values of $\left<k\right>$. The fraction of the removed nodes is denoted by $f$. For each strategy, $\lambda_2$ increases slightly in the early stages (\textit{i.e.}, $0<f<0.05$). Then, $\lambda_2$ starts to decrease even for the optimal sequential and perturbative strategies, which are designed to maximize $\lambda_2$. 
Surprisingly, in the network with larger $\left<k\right>$ (\FIG\ref{fig:1}(a)(2)), the optimal sequential strategy is not as efficient as the other strategies as $f$ increases. 
This is possible because the optimal sequential strategy finds the stepwise best strategy and does not take into account the performance after multiple nodes are removed. The perturbative strategy remains more efficient or as efficient as the optimal sequential strategy when $f$ is large.

For the WS model with different mean degrees (\FIG\ref{fig:1}(b)(1) and 1(b)(2)), the optimal sequential and perturbative strategies outperform the heuristic degree-based and betweenness-based strategies. As in the case of the ER random graph, the perturbative strategy is as efficient as the optimal sequential strategy.

The performances of the perturbative strategies are also good among the competitive strategies for different scale-free network models (\FIG\ref{fig:1}(c)(1), 1(c)(2), 1(d)(1), and 1(d)(2)). In Goh model(\FIG\ref{fig:1}(e)(1) and 1(e)(2)), thought it is not better than the degree-based strategy, the perturbative strategy is better than the optimal sequential strategy except for in the early stage (\textit{i.e.} $0<f<0.3$) in the Goh model with the smaller degree (\FIG\ref{fig:1}(e)(1)). Note that for the three scale-free network models, $\lambda_2$ continues to increase even until half the nodes are removed.

Changes in $\left<k\right>$ and the standard deviation of the degree with node removal according to the perturbative strategy are shown in \FIG\ref{fig:1}(a)(3), 1(a)(4), 1(b)(3), 1(b)(4), 1(c)(3), 1(c)(4), 1(d)(3), 1(d)(4), 1(e)(3), and 1(e)(4). The direction of changes in $\left<k\right>$ and that of the standard deviation of the degree depend on the network model. For example, in the Goh model, nodes with small degree are preferentially removed in general, especially for small $f$ (\FIG\ref{fig:1}(e)(3)). However, the removed nodes are not generally those with the smallest degrees; the degree-based strategy performs relatively poorly (\FIG\ref{fig:1}(b)(1), 1(b)(2), 1(c)(1), 1(c)(2), 1(d)(1), 1(d)(2), 1(e)(1), and 1(e)(2)). In contrast, in the ER, WS, BA, and HK models, the perturbative strategy removes nodes with appropriately large degree (\FIG\ref{fig:1}(a)(3), 1(b)(3), 1(c)(3), and 1(d)(3)). Similarly, the perturbative strategy increases $\lambda_2$ by increasing the heterogeneity of degree in the WS model (\FIG\ref{fig:1}(b)(4)) and by decreasing the same heterogeneity in the other four network models (\FIG\ref{fig:1}(a)(4), 1(c)(4), 1(d)(4), and 1(e)(4)). These show that the perturbative strategy adapts itself for each network.

\begin{figure}
\begin{center}
\includegraphics[width=15cm]{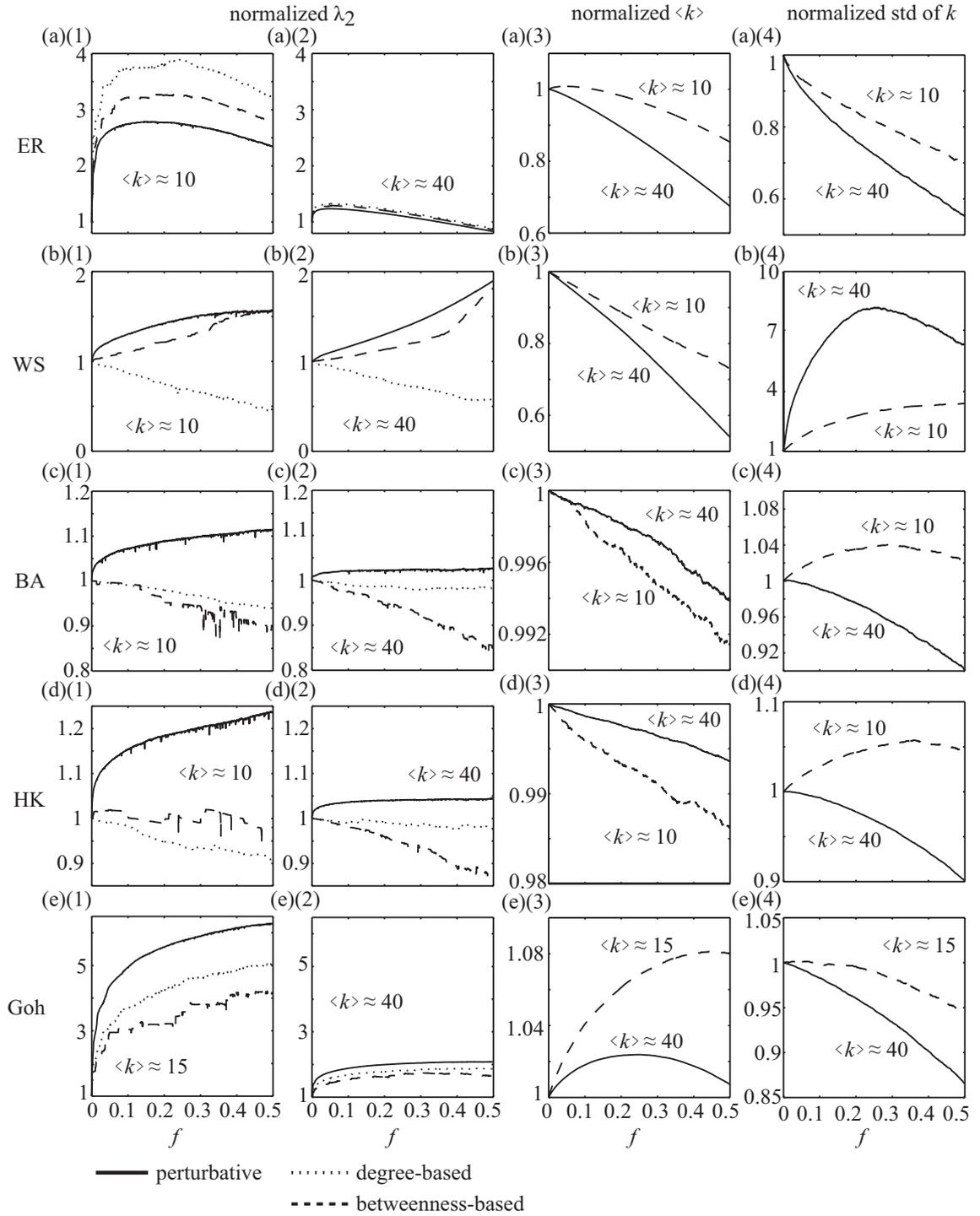}
\caption{Numerical results for (a) ER model, (b) WS model, (c) BA model, (d) HK model,  and (e) Goh model with $N=2000$.
We set $\left<k\right>=10$ and $\left<k\right>=40$ for each network model with one exception. We use $\left<k\right>=15$ instead of $\left<k\right>=10$ for the Goh model because the Goh model with $N=2000$, 
$\left<k\right>=10$, and $w_i= i^{-0.5}$ ($1\le i\le N$)
rarely yields a connected network.
}
\label{fig:2}
\end{center}
\end{figure}

Next, we compare the efficiency of different strategies on larger networks ($N=2000$). We exclude the optimal sequential strategy because the large $N$ hinders its implementation. In this set of numerical simulations, we are mainly concerned with the performance of the perturbative strategy.  The numerical results obtained on the basis of 5 realizations of each network are shown in \FIG\ref{fig:2}. These results are qualitatively the same as those obtained for the smaller networks shown in \FIG\ref{fig:1}. The perturbative strategy enhances $\lambda_2$
more efficiently than the other heuristic strategies except in ER models. In addition, the behavior of the perturbative strategy cannot be simply captured by the changes in $\left<k\right>$ or the standard deviation of the degree, which is again qualitatively the same as the results shown in \FIG\ref{fig:1}.

\subsection{Real networks}

We apply the proposed strategies to the largest connected component of the following real networks: the \textit{C.~elegans} neural network \cite{Chen06pnas,wormatlas}, \textit{E.~coli} metabolic network \cite{Jeong00nat}, e-mail social network \cite{Guimera03pre}, and macaque cortical network \cite{Felleman91cc-Young93pbs-SpornsZwi04ni}. We ignore the direction of links in the \textit{C.~elegans} neural network and the macaque cortical network, both of which are originally directed networks. In the \textit{C.~elegans} neural network, two neurons are regarded to be connected when they are connected by at least one chemical synapse or gap junction.

\begin{figure}
\begin{center}
\includegraphics[width=9cm]{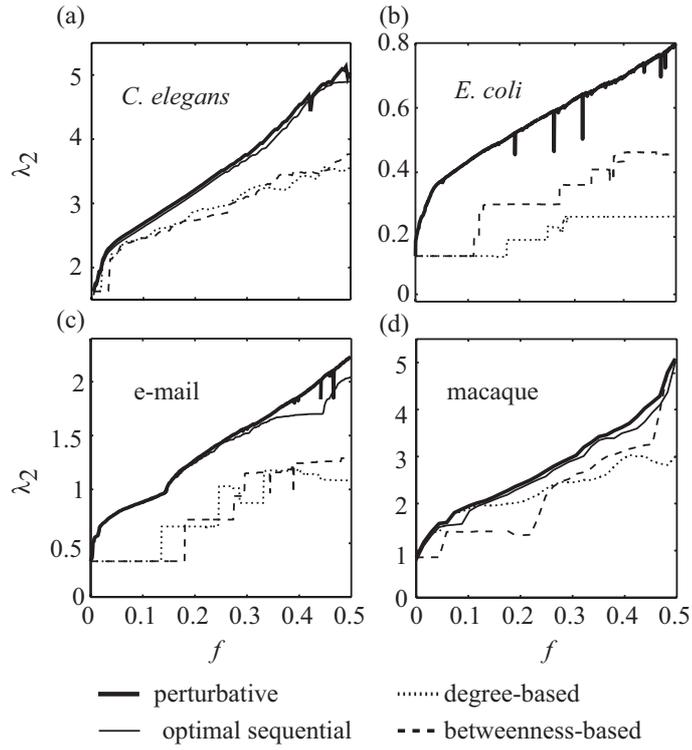}
\caption{Numerical results for real networks.
(a) \textit{C.~elegans} neural network with
$N=279$ and $\left<k\right>= 16.4$. (b)
\textit{E.~coli} metabolic network with $N=2268$ and
$\left<k\right>=4.96$. (c) E-mail social network with
$N=1133$ and $\left<k\right>=9.62$.
(d) Macaque cortical network with $N=71$ and
$\left<k\right>=12.3$.}
\label{fig:3}
\end{center}
\end{figure}

The efficiency of different strategies on these real networks is shown in \FIG\ref{fig:3}.  The perturbative strategy enhances $\lambda_2$ more efficiently in all the tested real networks than the degree-based and betweenness-based strategies. Except in the case of the \textit{E.~coli} metabolic network, which is too large for the optimal sequential strategy, the results for the optimal sequential strategy are shown as well (\FIG\ref{fig:3}(a), \ref{fig:3}(c), and \ref{fig:3}(d)).  The perturbative strategy performs roughly as well as the optimal sequential strategy in these networks.

\subsection{Comparison to the rewiring strategy}

One can alternatively enhance $\lambda_2$ by
rewiring links \cite{Donetti05prl,Donetti06jstatm}.
In the rewiring strategy, we sequentially rewire links to increase
$\lambda_2$. In each step, we
examine the increase in $\lambda_2$ for
all the possible patterns of single-link rewiring and adopt
the one that increases $\lambda_2$
by the largest amount.
To compare the performance of the node removal and the rewiring, we carry out numerical simulations
using the ER and BA models with $N=50$ and $\left<k\right> = 4$.
We simulate the rewiring process just once for each network
because the rewiring strategy is computationally costly.

\begin{figure}
\begin{center}
\includegraphics[width=9cm]{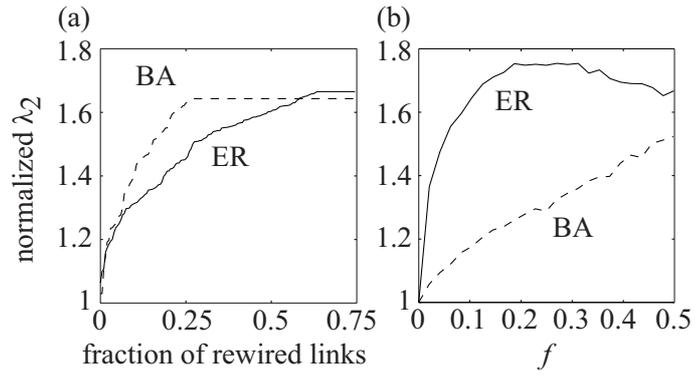}
\caption{Numerical results for the
ER and BA models with $N=50$.
(a) Results for link rewiring. The normalized $\lambda_2$ is plotted
against the number of rewired links.
(b) Results for node removal. The normalized $\lambda_2$ 
averaged over 10 realizations is plotted
against the number of removed nodes. We set $\left<k\right>= 4$
in both networks. 
}
\label{fig:4}
\end{center}
\end{figure}

The change in $\lambda_2$ relative to the initial value
during the rewiring process is shown in Fig.~\ref{fig:4}(a).
$\lambda_2$ is enhanced up to about 1.7 fold for both networks.
The corresponding
results for the sequential node removal according to the perturbative
strategy are shown in Fig.~\ref{fig:4}(b).
Roughly speaking, the performance of the perturbative strategy
is comparable to that of the rewiring strategy.
The perturbative strategy is superior to the rewiring strategy
for the ER model and vice versa for the BA model.
Because the rewiring strategy is computationally costly and may
be too demanding to be implemented
in some real applications,
the node removal according to the
perturbative strategy seems to be a feasible
choice for enhancing $\lambda_2$.

\subsection{Accuracy of the perturbative strategy}\label{sub:accuracy test}

When deriving
the perturbative strategy, we crucially
assumed that $\delta \bm u$ is negligible as compared to
$\bm u -u_i \hat{\bm e}_i$.
We justify this assumption as follows.
A node that is removed according to the perturbative strategy
tends to have large $\Delta \lambda_2$ 
and a small degree.
If the removed node has a small degree,
the number of the nonzero entries of the corresponding $\Delta
L$ is relatively small. Therefore, a relatively small number of the
entries of $\bm u$ would be directly affected by the node removal, and
we would obtain a small $\delta \bm u$.

To probe the validity of this assumption and quantify
the error in estimating $\Delta\lambda_2$,
we measure two kinds of relative estimation error
during the course of sequential node removal according to the perturbative
strategy.
The first quantity is 
the average of $|(\delta \bm u)_j/(\bm u-u_i\hat{\bm e}_i)_j|$
over node $j$ ($1\le j\le N$, $j\neq i$),
where $i$ is the index of the removed node.
The second quantity
is the difference between 
$\Delta\lambda_2$ obtained from the perturbative strategy
and the actual $\Delta\lambda_2$, which is normalized by the actual
$\Delta\lambda_2$. 
We take averages of these quantities over 200 generated networks
having $N=250$.

\begin{figure}
\begin{center}
\includegraphics[width=9cm]{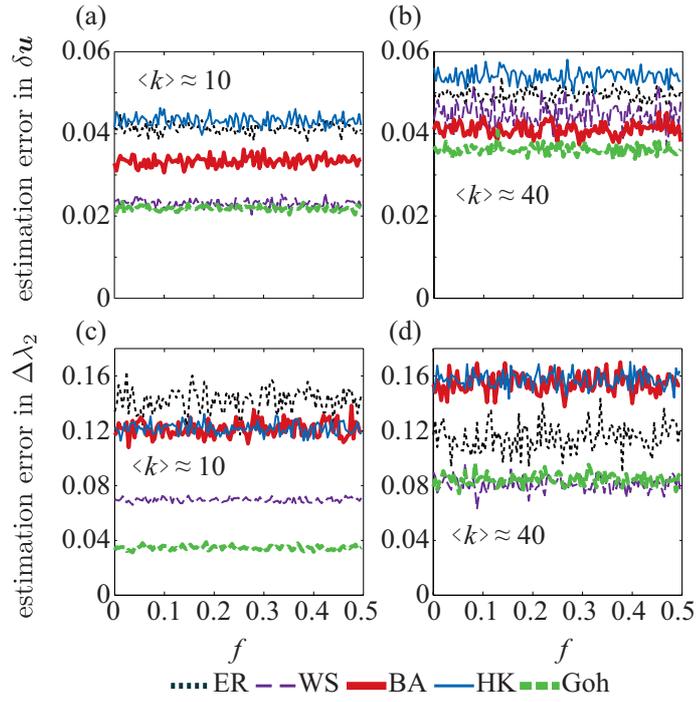}
\caption{(Color online) Relative estimation error in (a, b) $\delta \bm u$ and
(c, d) $\Delta\lambda_2$. See \SEC\ref{sub:accuracy test} for the
definitions of the relative estimation error.
We set (a, c) $\left<k\right>=10$ and (b, d) $\left<k\right>=40$.
We use $\left<k\right>=15$ instead of $\left<k\right>=10$ for the Goh model because the Goh model with $N=250$, 
$\left<k\right>=10$, and $w_i= i^{-0.5}$ ($1\le i\le N$)
rarely yields a connected network.}
\label{fig:5}
\end{center}
\end{figure}

For the five network models,
$|(\delta \bm u)_j/(\bm u-u_i\hat{\bm e}_i)_j|$
for
$\left<k\right>=10$ and $\left<k\right>=40$
is shown in Fig.~\ref{fig:5}(a) and \ref{fig:5}(b),
respectively.
The magnitude of $\delta \bm u$ relative to that of $\bm u-u_i\hat{\bm
e}_i$ is sufficiently small.
The relative estimation error in $\Delta\lambda_2$ 
for the removed nodes
is shown for $\left<k\right>=10$ and $\left<k\right>=40$
in Fig.~\ref{fig:5}(c) and
\ref{fig:5}(d), respectively.
As expected, the relative estimation error in $\Delta\lambda_2$ 
is generally small.
We conclude that, up to our numerical efforts,
the perturbative strategy does not suffer from a crucially large error.

\section{Conclusions}

We explored efficient strategies to sequentially remove nodes of networks in order to increase or maintain a large value of the spectral gap (\textit{i.e.}, second smallest eigenvalue of the Laplacian matrix) of the undirected and unweighted network. We introduced a perturbative strategy among others.  For a variety of networks, this strategy generally performed well as compared to heuristic strategies in which we sequentially remove the nodes with the smallest degree or the smallest betweenness centrality. In most of our numerical results, the spectral gap increased until the removal of a fairly large fraction of nodes ($\approx$ 50\%). Occasionally, the perturbative strategy is even more efficient than the optimal sequential strategy, despite its decreased computational cost. Although we focused on unweighted networks, the extension of the perturbative strategy to the case of weighted networks is straightforward.

In chaotic dynamical systems on networks, synchronization is often
facilitated by a small value of $R = \lambda_N/\lambda_2$, where
$\lambda_2$ and $\lambda_N$ are the second smallest eigenvalue and the
largest eigenvalue of the Laplacian matrix, respectively. Dynamics
whose synchronizability is determined by $R$ belongs to the class III
\cite{Boccaletti06PR,Almendral09pre} (also termed type I \cite{Arenas08pr}).
In contrast, we have been concerned
with the synchronization of the class II
\cite{Boccaletti06PR,Almendral09pre} (also termed type II \cite{Arenas08pr})
dynamics in which synchronization is facilitated in networks with
large $\lambda_2$. To address class III or type I
synchronizability, we developed the perturbative strategy for
minimizing $R$ upon the removal of nodes and assessed its efficiency
on some model networks. However, the results were generally poor
(results not shown). The perturbative strategy failed mainly because
it does not accurately estimate the change in $\lambda_N$.
The applicability of our results is limited to class II or type II dynamics.

\begin{acknowledgements}
We thank Hiroshi Kori and Ralf T\"{o}njes for their valuable discussions. N.M. acknowledges the support through the Grants-in-Aid for Scientific Research (Nos. 20760258 and 20540382, and Innovative Areas ``Systems Molecular Ethology'') from the Ministry of Education, Culture, Sports, Science and Technology (MEXT), Japan. T.W. acknowledges the support from the Japan Society for the Promotion of Science (JSPS) Research Fellowship for Young Scientists (222882).
\end{acknowledgements}

\clearpage
\newpage

\pagestyle{empty}

\clearpage

\clearpage

\clearpage

\clearpage

\end{document}